\documentclass[12pt]{article}
\usepackage{graphics} 
\usepackage{cite}
\usepackage{epsfig}
\usepackage{amssymb}
\newcommand  {\etal}     {{\it et al.}}

\newcommand  {\APC}      {{\it Adv. Protein Chem.\ }}

\newcommand  {\Bioch}    {{\it Biochemistry\ }}

\newcommand  {\BJ}       {{\it Biophys.~J.\ }}

\newcommand  {\COSB}     {{\it Curr.\ Opin.\ Struct.\ Biol.\ }}

\newcommand  {\EL}       {{\it Europhys.\ Lett.\ }}

\newcommand  {\JACS}     {{\it J.\ Am.\ Chem.\ Soc.\ }}
\newcommand  {\JBC}      {{\it J.\ Biol.\ Chem.\ \ }}

\newcommand  {\JCP}      {{\it J.\ Chem.\ Phys.\ }}
\newcommand  {\JMB}      {{\it J.\ Mol.\ Biol.\ }}

\newcommand  {\Nat}      {{\it Nature\ }}
\newcommand  {\NSB}      {{\it Nat.\ Struct.\ Biol.\ }}

\newcommand  {\Pro}      {{\it Proteins\ }}

\newcommand  {\ProSci}   {{\it Protein\ Sci.\ }}

\newcommand  {\PNAS}     {{\it Proc.\ Natl.\ Acad.\ Sci.\ USA\ }}

\newcommand  {\PRL}      {{\it Phys.\ Rev.\ Lett.\ }}

\newcommand  {\Str}      {{\it Structure\ }}
\newcommand  {\TBS}      {{\it Trends Biochem. Sci.\ }}

\newcommand{\beq}{\begin{equation}}
\newcommand{\eeq}{\end{equation}}
\newcommand{\beqa}{\begin{eqnarray}}
\newcommand{\eeqa}{\end{eqnarray}}
\newcommand{\bea}{\begin{eqnarray}}
\newcommand{\eea}{\end{eqnarray}}
\newcommand   {\ev}[1]   {\langle #1\rangle}
\newcommand   {\Ab}      {A$\beta$} 
\newcommand   {\Aba}     {A$\beta_{16-22}$} 
\newcommand   {\Abb}     {A$\beta_{16-35}$} 
\newcommand   {\Abc}     {A$\beta_{10-35}$} 
\newcommand   {\Abd}     {A$\beta_{1-40}$} 
\newcommand   {\Abe}     {A$\beta_{1-42}$} 
\newcommand   {\Abf}     {A$\beta_{11-25}$} 
\newcommand   {\Abg}     {A$\beta_{34-42}$} 
\newcommand   {\Abx}     {A$\beta_{16-20}$}
\newcommand   {\Ca}      {C${}_{\alpha}$}
\newcommand   {\Cb}      {C${}_{\beta}$}
\newcommand   {\Cd}      {C${}_{\delta}$}
\newcommand   {\Cg}      {C${}_{\gamma}$}
\newcommand   {\Cp}      {C${}^{\prime}$}
\newcommand   {\rc}      {r^{\mbox{{\scriptsize c}}}}

\newcommand   {\Eloc}    {E_{\mbox{{\scriptsize loc}}}}
\newcommand   {\Eev}     {E_{\mbox{{\scriptsize ev}}}}
\newcommand   {\Ehb}     {E_{\mbox{{\scriptsize hb}}}}

\newcommand   {\Ehp}     {E_{\mbox{{\scriptsize hp}}}}
\newcommand   {\Mij}     {M_{IJ}}

\newcommand   {\Nc}      {N_{\mbox{{\scriptsize c}}}}
\newcommand   {\Cv}      {C_{\mbox{{\scriptsize v}}}}

\newcommand   {\kev}     {\kappa_{\mbox{{\scriptsize ev}}}}
\newcommand   {\kloc}    {\kappa_{\mbox{{\scriptsize loc}}}}

\newcommand   {\ehba}    {\epsilon^{(1)}_{\mbox{{\scriptsize hb}}}}
\newcommand   {\ehbb}    {\epsilon^{(2)}_{\mbox{{\scriptsize hb}}}}

\newcommand   {\shb}     {\sigma_{\mbox{{\scriptsize hb}}}}

\newcommand   {\np}      {n_{+}}
\newcommand   {\nm}      {n_{-}}

\begin{document}

\begin{flushright}
LU TP 04-18\\
August 5, 2004
\end{flushright}

\vspace{0.4in}

\begin{center}

{\LARGE \bf Oligomerization of Amyloid \Aba}\\ 
\vspace{6pt}  
{\LARGE \bf Peptides Using Hydrogen Bonds}\\
\vspace{6pt}  
{\LARGE \bf and Hydrophobicity Forces}\\

\vspace{.6in}

\large
Giorgio Favrin, Anders Irb\"ack and Sandipan 
Mohanty\footnote{E-mail: favrin,\,anders,\,sandipan@thep.lu.se}\\   
\vspace{0.10in}
Complex Systems Division, Department of Theoretical Physics\\ 
Lund University,  S\"olvegatan 14A,  SE-223 62 Lund, Sweden \\
{\tt http://www.thep.lu.se/complex/}\\

\vspace{0.3in}  

Submitted to \BJ

\end{center}
\vspace{0.3in}
\normalsize
Abstract:\\
The 16--22 amino acid fragment of the $\beta$-amyloid peptide associated
with the Alz-heimer's disease, \Ab, is capable of forming amyloid fibrils. 
Here we study the aggregation mechanism of \Aba\ peptides
by unbiased thermodynamic simulations at the atomic level for systems
of one, three and six \Aba\ peptides. We find that the isolated \Aba\ peptide 
is mainly a random coil in the sense that both the $\alpha$-helix and 
$\beta$-strand contents are low, whereas the three- and six-chain 
systems form aggregated structures with a high $\beta$-sheet content. 
Furthermore, in agreement with experiments on \Aba\ fibrils, we 
find that large parallel $\beta$-sheets are unlikely to form. For the 
six-chain system, the aggregated structures can have many different shapes, 
but certain particularly stable shapes can be identified.    

\vspace{24pt}

\newpage 

\section{Introduction}

The fibrillar aggregates that characterize amyloid diseases, such as 
the Alzheimer's disease, are formed by specific peptides or proteins. However, 
it is known that several non-disease-related proteins are capable 
of forming similar amyloid structures~\cite{Rochet:00,Dobson:03}, and
that the aggregation of such proteins can be cytotoxic~\cite{Bucciantini:02}. 
This suggests, first, that 
polypeptide chains have a general tendency to form amyloid structures and, 
second, that natural proteins should have evolved mechanisms to avoid this 
tendency. Such mechanisms have indeed been 
proposed~\cite{Otzen:00,Broome:00,Richardson:02}.
The propensity of a given polypeptide chain to form amyloid fibrils 
depends, nevertheless, on its amino acid 
sequence~\cite{West:99,Villegas:00,Hammarstrom:02,LopezdelaPaz:02,Chiti:03},
and short sequence stretches promoting amyloid formation have been  
identified~\cite{LopezdelaPaz:04,Ventura:04}.  

While the structure of amyloid fibrils is not known in atomic detail, 
there is ample evidence from X-ray fiber diffraction studies that
the core of the typical amyloid fibril is composed of $\beta$-sheets 
whose strands run perpendicular to the fibril axis~\cite{Sunde:97}. 
More detailed information is available, for example, for fibrils made from  
different fragments of the Alzheimer's \Ab\ peptide. In particular,
there is evidence from solid-state NMR studies for a parallel 
organization of the $\beta$-strands in  \Abc~\cite{Burkoth:00} and 
\Abd~\cite{Petkova:02} fibrils, and for an  
antiparallel organization in \Abg~\cite{Lansbury:95}, \Abf~\cite{Petkova:04} 
and \Aba\ fibrils~\cite{Balbach:00,Gordon:04}. 
Most of these fragments contain the hydrophobic
\Abx\ segment (KLVFF), which is known to be important in the 
\Ab-\Ab\ interaction~\cite{Tjernberg:96}.

Small peptides like \Aba\ are well suited as model systems for probing 
the mechanisms of aggregation and fibril formation, and are being studied
not only {\it in vitro} but also {\it in silico}. Computer simulations of 
simplified~\cite{Bratko:01,Harrison:01,Dima:02,Jang:04,
Friedel:04} and atomic~\cite{Ma:02a,Ma:02b,Klimov:03,Gsponer:03,Paci:04} 
models have provided useful insights into the aggregation behavior of some 
peptide systems. To properly explore the free-energy landscape of aggregation 
at the atomic level is, nevertheless, a computational challenge.      

Here we investigate the formation and properties of \Aba\ oligomers  
by unbiased Monte Carlo (MC) simulations of systems with up to six chains,
using a sequence-based atomic model with an effective potential based
on hydrogen bonds and hydrophobic attraction (no explicit water molecules).  
The same model has previously 
been used to study the folding of individual 
peptides~\cite{Irback:03,Irback:04a,Irback:04b}. It was shown 
that this model is able to fold several different peptides, 
both $\alpha$-helical and $\beta$-sheet peptides, for one and 
the same choice of parameters. The calculated melting behaviors 
were, moreover, in good agreement with experimental data for all 
these peptides.  
 
\newpage 
   
\section{Model and Methods}\label{MM}

The main object of study in this paper is the peptide
\Aba, given by 
acetyl-Lys-Leu-Val-Phe-Phe-Ala-Glu-NH${}_2$. 
We consider systems of one, three and six \Aba\ peptides. 
The multichain systems are contained in periodic boxes. 
All the interactions are short range, which makes the implementation
of the periodic boundary conditions straightforward. 
The box sizes are (35\AA)${}^3$ and  (44\AA)${}^3$  
for three and six chains, 
respectively, corresponding to a constant peptide concentration. 
For computational efficiency, the peptide concentration is taken
to be high.  

Our model~\cite{Irback:03,Irback:04a,Irback:04b} contains all atoms 
of the peptide chains, including hydrogen atoms.
The model assumes fixed bond lengths, bond angles and peptide 
torsion angles ($180^\circ$),  
so that each amino acid only has the Ramachandran torsion 
angles $\phi$, $\psi$ and a number of side-chain torsion angles as its 
degrees of freedom. Numerical values of the geometrical parameters 
held constant can be found elsewhere~\cite{Irback:03}.

The interaction potential 
\begin{equation}
  E=\Eev+\Eloc+\Ehb+\Ehp
\label{energy}
\end{equation}
is composed of four terms, which we describe next. 
Energy parameters 
are given on a scale such that a temperature of $T=300$\,K 
corresponds to $kT\approx 0.447$ 
($k$ is Boltzmann's constant)~\cite{Irback:04b}. 

The first term in Eq.~\ref{energy}, $\Eev$, represents excluded-volume effects
and has the form
\beq
\Eev=\kev \sum_{i<j}
\biggl[\frac{\lambda_{ij}(\sigma_i+\sigma_j)}{r_{ij}}\biggr]^{12}\,,
\label{ev}\eeq
where the summation is over pairs of atoms $(i,j)$, $\kev=0.10$,
and $\sigma_i=1.77$, 1.75, 1.55, 1.42 and 1.00\,\AA\ for
S, C, N, O and H atoms, respectively. The parameter $\lambda_{ij}$
has the value 0.75 for all pairs except those connected by
three covalent bonds, for which $\lambda_{ij}=1$.
When the two atoms belong to different chains, we always use 
$\lambda_{ij}=0.75$. To speed up the calculations, Eq.~\ref{ev} is
evaluated using a cutoff of $\rc_{ij}=4.3\lambda_{ij}$\,\AA, and
pairs with fixed separation are omitted. 

The second energy term, $\Eloc$, is a local intrachain potential. 
It has the form 
\beq
\Eloc=\kloc \sum_I
\left(\sum \frac{q_iq_j}{r_{ij}^{(I)}/{\rm \AA}}\right)\,,
\label{loc}\eeq
where the inner sum represents the interactions between
the partial charges of the backbone NH and \Cp O groups in one
amino acid, $I$. This potential is not used for Gly and Pro amino acids 
which have very different $\phi$, $\psi$ distributions, but is the same 
for all other amino acids.     
The inner sum has four terms (N\Cp, NO, H\Cp\ and HO) 
which depend only on the $\phi$ and $\psi$ angles for amino acid $I$. 
The partial charges are taken as $q_i=\pm 0.20$ for H and N 
and $q_i=\pm 0.42$ for \Cp\ and O~\cite{Branden:91}, and we put $\kloc=100$, 
corresponding to a dielectric constant of $\epsilon_r\approx 2.5$.

The third term of the energy function is the hydrogen-bond energy $\Ehb$, 
which has the form
\begin{equation}
  \Ehb= \ehba \sum_{{\rm bb-bb}}
  u(r_{ij})v(\alpha_{ij},\beta_{ij}) +
  \ehbb \sum_{{\rm sc-bb}}
  u(r_{ij})v(\alpha_{ij},\beta_{ij})\,,
  \label{hbonds}
\end{equation}
where the two functions $u(r)$ and $v(\alpha,\beta)$ are given by
\begin{eqnarray}
u(r)&=& 5\bigg(\frac{\shb}{r}\bigg)^{12} -
             6\bigg(\frac{\shb}{r}\bigg)^{10}\label{u}\\
v(\alpha,\beta)&=&\left\{
        \begin{array}{ll}
             (\cos\alpha\cos\beta)^{1/2} &
              \ {\rm if}\ \alpha,\beta>90^{\circ}\label{v}\\
             0  & \ \mbox{otherwise}
        \end{array} \right.
\end{eqnarray}
We consider only hydrogen bonds between NH and CO groups, and
$r_{ij}$ denotes the HO distance, $\alpha_{ij}$ the NHO angle and
$\beta_{ij}$ the HOC angle. The parameters $\ehba$, $\ehbb$ and
$\shb$ are taken as 3.1, 2.0 and 2.0\,\AA, respectively. 
The function $u(r)$ is calculated using a cutoff of $\rc=4.5$\,\AA.
The first sum in Eq.~\ref{hbonds} contains backbone-backbone
interactions, while the second sum contains interactions between
charged side chains (Asp, Glu, Lys and Arg) and the backbones.
For intrachain hydrogen bonds we make two restrictions.
First, we disallow backbone NH (\Cp O) groups to make hydrogen 
bonds with the two nearest backbone \Cp O (NH) groups on each side of them. 
Second, we forbid hydrogen bonds between the side chain of one amino acid  
with the nearest donor or acceptor on either side of its \Ca. 
For interchain hydrogen bonds, we make no such restrictions. As a
simple form of context dependence, we assign a reduced strength to 
hydrogen bonds involving chain ends, which tend to be exposed to water.   
Following the experimental studies of the \Aba\ 
peptide~\cite{Balbach:00,Gordon:04}, we have  
used acetyl and amide capping groups at the ends. A 
hydrogen bond involving one or two such groups is reduced in strength  
by factors of 2 and 4, respectively.   

The fourth energy term, $\Ehp$, represents an effective hydrophobic attraction
between nonpolar side chains. It has the pair-wise additive form
\beq
\Ehp=-\sum_{I<J}\Mij C_{IJ}\,,
\label{hp}\eeq
where $C_{IJ}$ is a measure of the degree of contact between
side chains $I$ and $J$, and $\Mij$ sets the energy that a pair
in full contact gets. The matrix $\Mij$ is defined in Table~\ref{tab:1}. 
To calculate $C_{IJ}$ we use a predetermined set of atoms, $A_I$, 
for each side chain $I$. We define $C_{IJ}$ as 
\beq
C_{IJ}=\frac{1}{N_I+N_J} \biggl[\,
\sum_{i\in A_I}f(\min_{j\in A_J} r_{ij}^2) +
\sum_{j\in A_J}f(\min_{i\in A_I} r_{ij}^2)
\,\biggr]\,,
\label{Rij}\eeq
where the function $f(x)$ is given by $f(x)=1$ if $x<A$, $f(x)=0$ if $x>B$,
and $f(x)=(B-x)/(B-A)$ if $A<x<B$ [$A=(3.5\,{\rm \AA}){}^2$ and
$B=(4.5\,{\rm \AA}){}^2$]. Roughly speaking, $C_{IJ}$ is the fraction 
of atoms in $A_I$ or $A_J$ that are in contact with some atom from 
the other side chain. For Pro, the set $A_I$ consists of the 
\Cb, \Cg\ and \Cd\ atoms. The definition of $A_I$ for the other 
hydrophobic side chains has been given elsewhere~\cite{Irback:03}. 
For pairs that are nearest or next-nearest neighbors along the same chain, 
we use a reduced strength for the hydrophobic attraction; 
$\Mij$ is reduced by a factor of 2 for next-nearest neighbors, 
and taken to be 0 for nearest neighbors. 

\begin{table}[t]
\begin{center}
\begin{tabular}{rlccc}
                           & & I   & II & III \\
\hline
I& Ala                       & 0.0   & 0.1  & 0.1   \\
II& Ile, Leu, Met, Pro, Val  &       & 0.9  & 2.8   \\
III& Phe, Trp, Tyr           &       &      & 3.2
\end{tabular}
\caption{
The hydrophobicity matrix $\Mij$. Hydrophobic amino acids are divided 
into three categories. The matrix $\Mij$ represents the size of 
hydrophobicity interaction when an amino acid of type $I$ is in contact
with an amino acid of type $J$.} 
\label{tab:1}
\end{center}
\end{table}

To study the thermodynamic behavior of this model, 
we use simulated tempering~\cite{Lyubartsev:92,Marinari:92,Irback:95}
in which the temperature is a dynamical variable. 
For a review of simulated tempering and other generalized-ensemble
techniques for protein folding, see Ref.~\cite{Hansmann:99}. We study 
the one- and three-chain systems at eight different temperatures, 
ranging from 275\,K to 369\,K, and the six-chain system at 
seven temperatures, ranging from 287\,K to 369\,K.

Our simulations are carried out using two different elementary moves 
for the backbone degrees of freedom: 
first, the highly non-local pivot move in which a single backbone 
torsion angle is turned; and second, a semi-local method~\cite{Favrin:01} that
works with up to eight adjacent backbone degrees of freedom, which are turned 
in a coordinated manner. Side-chain angles are updated one by one.  
In addition to these updates, we also use 
rigid-body translations and rotations of whole chains. 
Every update involves a Metropolis accept/reject step, 
thus ensuring detailed balance. All our simulations are started from 
random configurations. All statistical errors quoted are 1$\sigma$ errors 
obtained from the variation between independent runs. We performed 
9 runs with $10^8$ elementary MC steps for $\Nc=1$, 
11 runs with $10^9$ MC steps for $\Nc=3$, and 18 runs
with $2\cdot10^9$ MC steps for $\Nc=6$. Each of the $\Nc=6$ runs
required about 12 CPU days on a 1.6\,GHz computer.   

To characterize the behavior of these systems, we first determine the
secondary structure. For a chain with $N$ amino acids, we define the 
$\alpha$-helix and $\beta$-strand contents as the fractions of 
the $N-2$ inner amino acids with their ($\phi,\psi$) pair in the 
$\alpha$-helix and $\beta$-strand regions of the Ramachandran space. 
We assume that $\alpha$-helix corresponds to 
$-90^\circ<\phi<-30^\circ$, $-77^\circ<\psi<-17^\circ$
and that $\beta$-strand corresponds to $-150^\circ<\phi<-90^\circ$,
$90^\circ<\psi<150^\circ$.
The average $\alpha$-helix and $\beta$-strand contents, over all the 
chains of the system, are denoted by $H$ and $S$, respectively.

\begin{figure}[t]
\begin{center}
\epsfig{figure=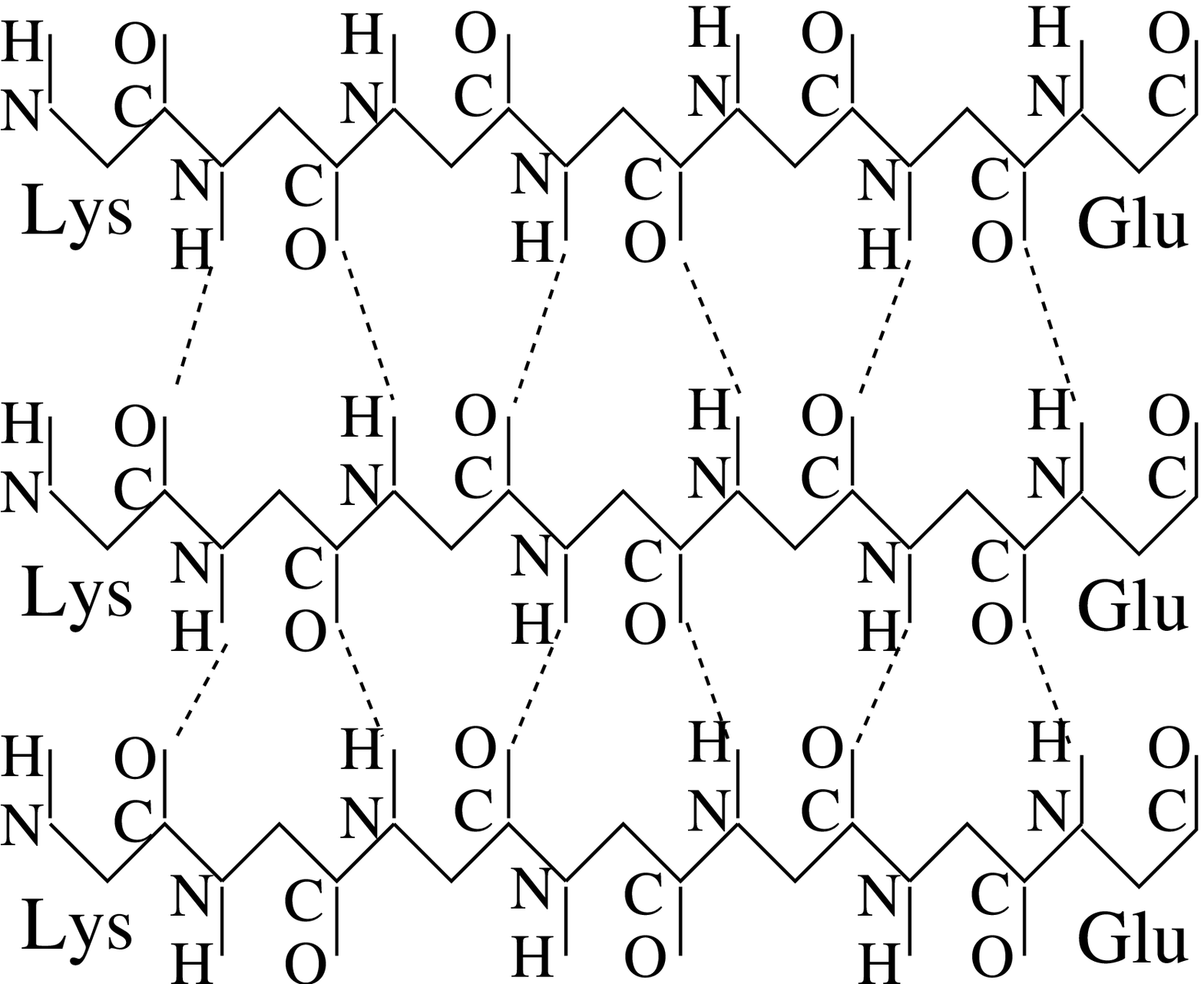,width=5.8cm}
\hspace{15mm}
\epsfig{figure=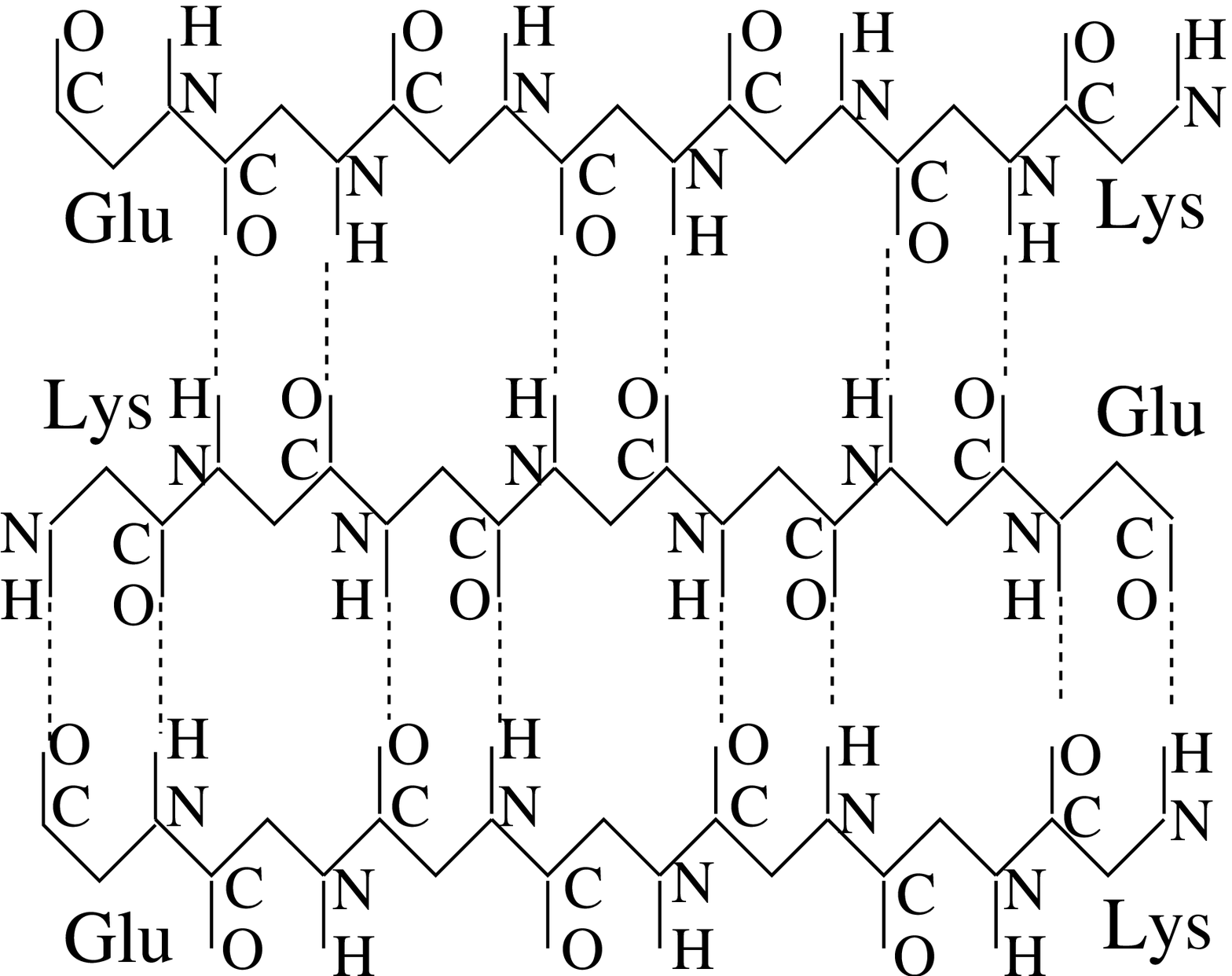,width=6cm}
\end{center}
\vspace{-3mm}
\caption{Schematic illustrations of the hydrogen bond patterns for  
in-register, parallel $\beta$-strands (left) and in-register,
antiparallel $\beta$-strands (right).} 
\label{fig:1}\end{figure}

To distinguish between parallel and antiparallel $\beta$-sheet structure,
we examine the orientation of end-to-end vectors. For a given multichain 
configuration, we first determine all pairs of chains 
such that (i)  
their interchain hydrogen bond energy is less than $-1.5\ehba$ 
(roughly corresponding to 2--3 hydrogen bonds), and (ii) both chains 
have a $\beta$-strand content
higher than 0.5.
For each such pair of chains, we then  
calculate the scalar product of their normalized end-to-end unit vectors. 
If this scalar product is greater than 0.7 (less than $-0.7$), 
we say that the two chains are parallel (antiparallel). 
We denote the numbers of parallel and antiparallel pairs of chains 
by $\np$ and $\nm$, respectively. Fig.~\ref{fig:1} illustrates       
the hydrogen-bond patterns in parallel and antiparallel 
$\beta$-sheets.

\newpage

\section{Results and Discussion}

\begin{figure}[t]
\begin{center}
\epsfig{figure=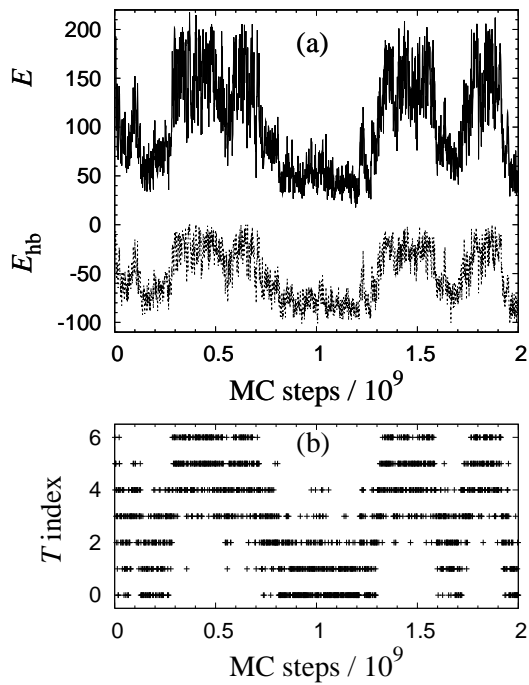,width=7cm}
\end{center}
\vspace{-3mm}
\caption{Monte Carlo evolution in a simulated-tempering run for 
$\Nc=6$ \Aba\ peptides.  
(a) The total energy $E$ (full line) and the hydrogen-bond energy $\Ehb$ 
(dashed line), both in kcal/mol. (b) The temperature index $k$. 
There are seven allowed temperatures $T_k$, satisfying 
$T_0=287$\,K${}<T_1<\ldots<T_6=369$\,K. Measurements are taken every
$10^6$ MC steps.}
\label{fig:2}
\end{figure}

Using the model described in the previous section,
we study the thermodynamics of systems of $\Nc$ \Aba\ peptides 
for $\Nc=1$, 3 and 6. Fig.~\ref{fig:2} 
illustrates the Monte Carlo evolution
in one of 18 independent simulated-tempering runs for the six-chain 
system. In the course of the run, aggregated low-energy structures form and
dissolve several times. 

\newpage

\subsection{Secondary Structure}

\begin{figure}[t]
\begin{center}
\epsfig{figure=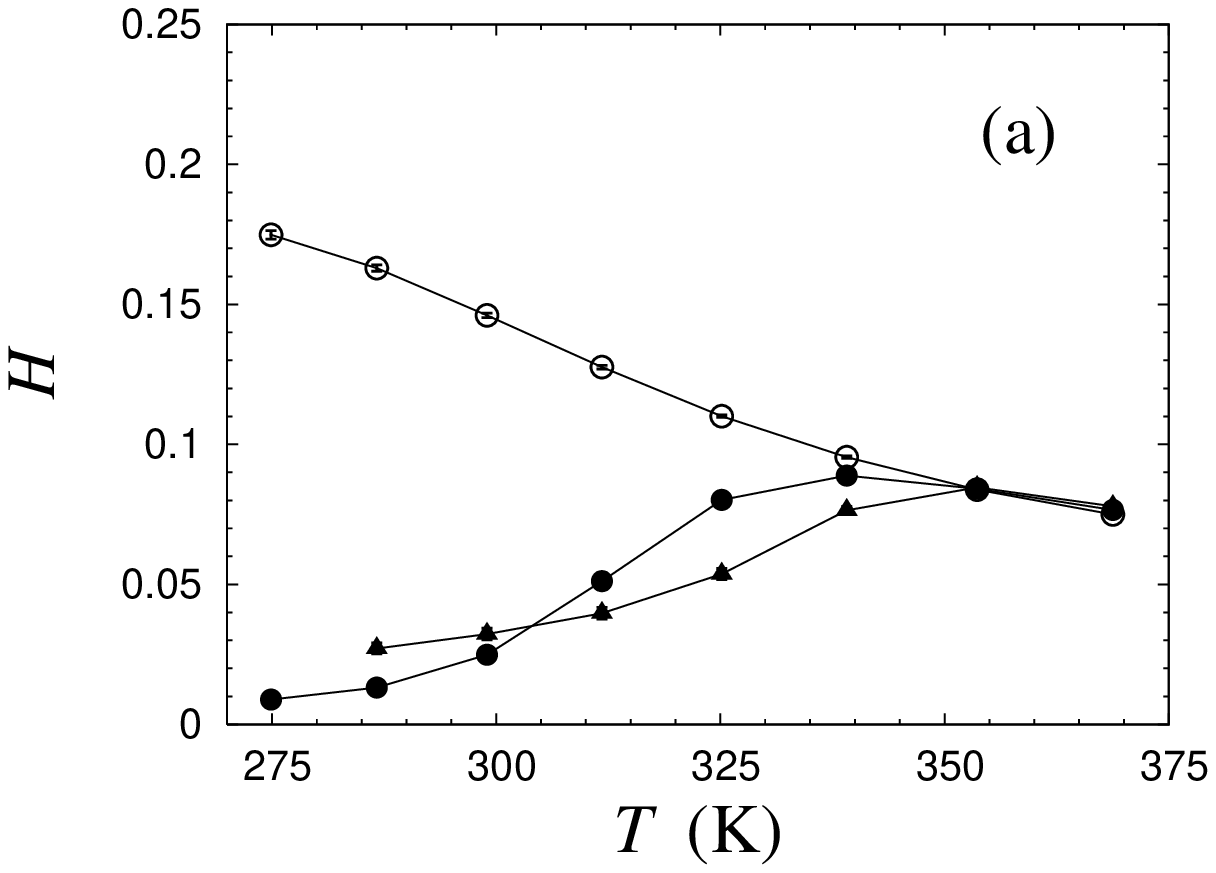,width=7cm}
\epsfig{figure=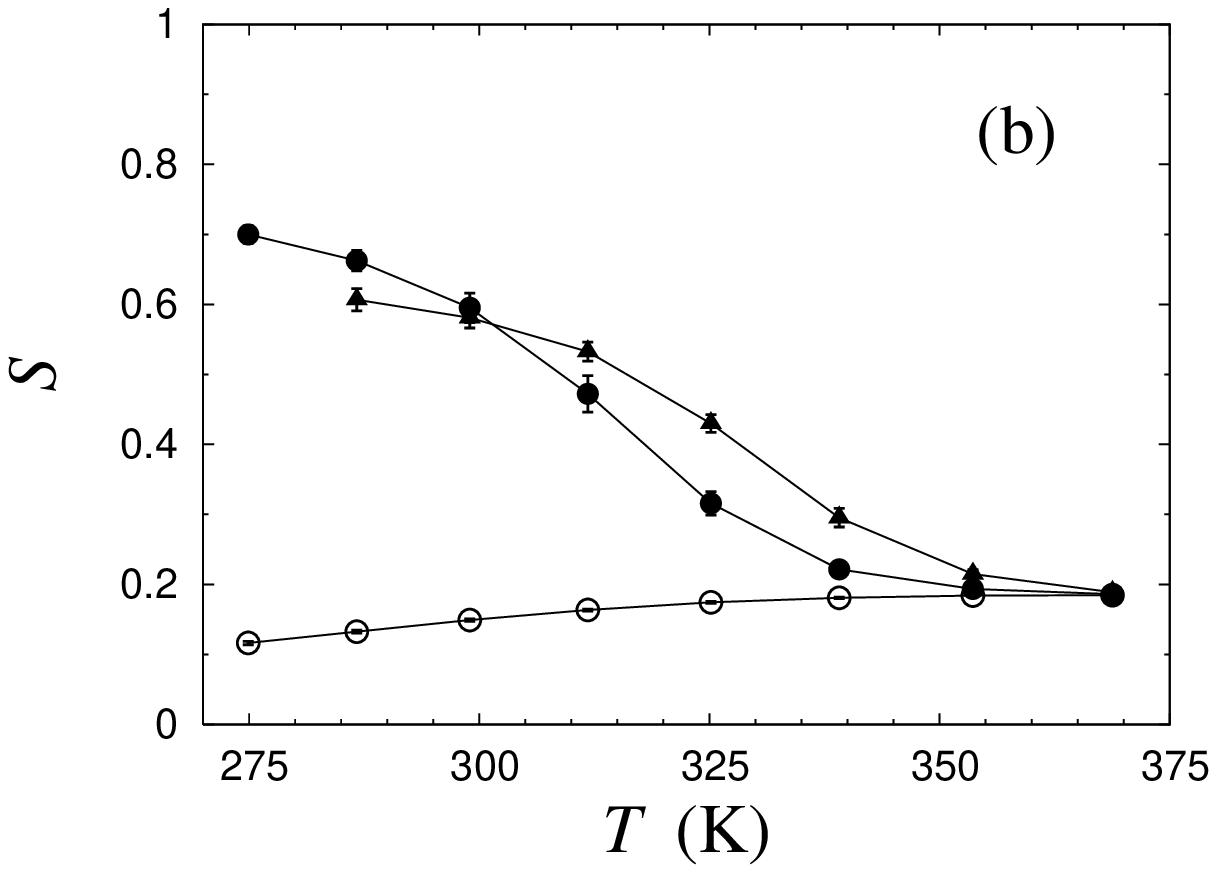,width=7cm}
\end{center}
\vspace{-3mm}
\caption{(a) The $\alpha$-helix content $H$ against temperature $T$
for \Aba\ for $\Nc=1$ ($\circ$), $\Nc=3$ ($\bullet$) and 
$\Nc=6$ ($\blacktriangle$). 
Lines joining data points are only a guide for the eye.  
(b) Same for the $\beta$-strand content $S$.
Note that the scales in (a) and (b) are different.}   
\label{fig:3}\end{figure}

Fig.~\ref{fig:3} shows the $\alpha$-helix and $\beta$-strand contents 
$H$ and $S$, as defined in the previous section, against temperature 
for different $\Nc$. For $\Nc=1$, we see that both $H$ and $S$ 
are small at all temperatures studied, although $H$ increases with 
decreasing temperature. So, in our model, the \Aba\ monomer is 
mainly a random coil throughout this temperature range. 
The $\Nc=3$ and $\Nc=6$ systems show a qualitatively different 
behavior; $S$ increases sharply with decreasing temperature, to values
of $S=0.6$ and higher, whereas $H$ is very small. These results 
clearly demonstrate 
that unless the temperature is too high, the three- and six-chain systems 
self-assemble into ordered structures with a high $\beta$-strand content.

\begin{figure}[t]
\begin{center}
\epsfig{figure=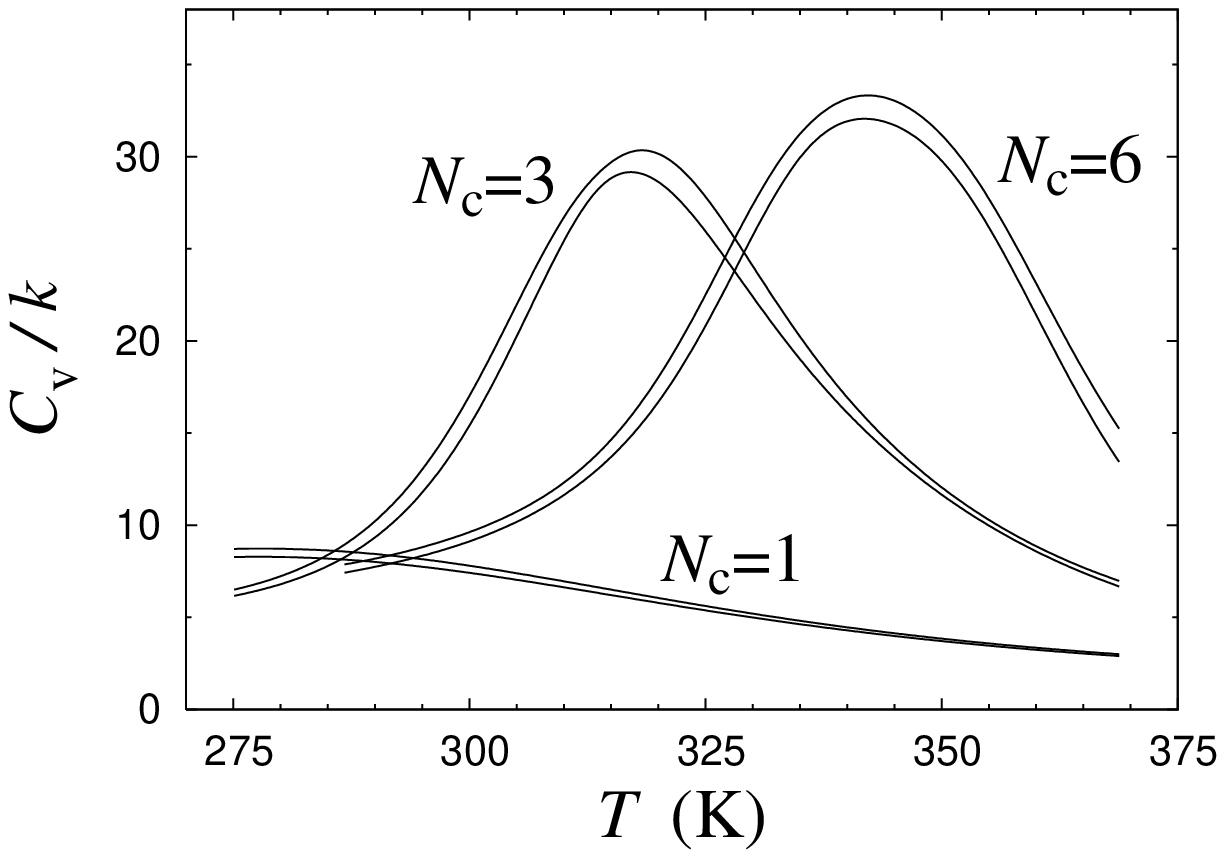,width=7cm}
\end{center}
\vspace{-3mm}
\caption{Specific heat $\Cv$ against 
temperature $T$  for $\Nc=1$, 3 and 6 \Aba\ peptides, as obtained 
using histogram reweighting techniques~\cite{Ferrenberg:88}. The
bands are centered around the expected values and show statistical
1$\sigma$ errors. $\Cv$ is defined as 
$\Cv=(\Nc N)^{-1}d\ev{E}/dT= (\Nc NkT^2)^{-1}(\ev{E^2}-\ev{E}^2)$,
where $\Nc$ is the number of chains, $N$ is the number of amino acids per 
chain, and $\ev{O}$ denotes a Boltzmann average of variable $O$.} 
\label{fig:4}\end{figure}

The temperature at which the aggregation sets in depends strongly  
on the peptide concentration, and exploring that dependence is beyond 
the scope of the present study. We note, however, that the $\beta$-sheet
formation sets in at a higher temperature for $\Nc=6$ than for $\Nc=3$. 
This fact is also reflected in the behavior of the specific heat, as
shown in Fig.~\ref{fig:4}. For $\Nc=3$ and $\Nc=6$, the specific heat $\Cv(T)$ 
exhibits a pronounced peak. As the system size increases from $\Nc=3$ to 
$\Nc=6$, the peak is shifted towards higher temperature. Near the peak, the 
energy distribution is broad (data not shown), showing that both 
aggregated low-energy and unstructured high-energy states occur with
a significant frequency at these temperatures. 
   
Our results for $\Nc=1$ and $\Nc=3$ can be compared with results from 
molecular dynamics simulations with 
explicit water by Klimov and Thirumalai~\cite{Klimov:03}. 
Using somewhat different definitions of $H$ and $S$ and  
a temperature of $T=300$\,K, these authors found that
$H=0.11$ and $S=0.33$ for $\Nc=1$, and $H=0.26$ and $S=0.30$ for $\Nc=3$. 
Our $\Nc=1$ results (see Fig.~\ref{fig:3})
are in reasonable agreement with theirs, 
given that we use a stricter definition 
and $\beta$-strands. However, our $\Nc=3$ results disagree 
with theirs. They obtained a smaller $\beta$-strand content and  
a larger $\alpha$-helix content compared to their own $\Nc=1$ results;
whereas we observe a much larger $\beta$-strand content for $\Nc=3$
compared to $\Nc=1$. 

\begin{figure}[t]
\begin{center}
\epsfig{figure=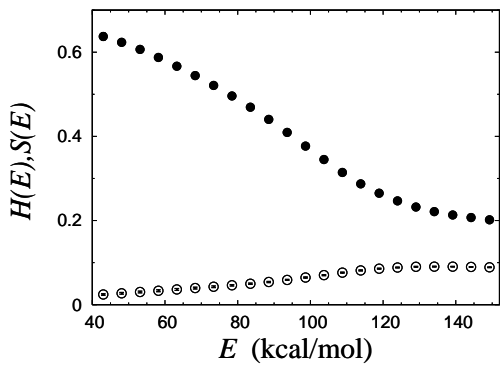,width=7cm}
\end{center}
\vspace{-3mm}
\caption{The $\alpha$-helix ($\circ$) and $\beta$-strand ($\bullet$) 
profiles $H(E)$ and $S(E)$ (see the text) for the six-chain \Aba\ 
system at $T=325$\,K.} 
\label{fig:5}
\end{figure}

For the $\Nc=3$ system, Klimov and Thirumalai~\cite{Klimov:03} 
furthermore found evidence for an obligatory $\alpha$-helical 
intermediate. To see whether or not such an intermediate exists 
in our model, we divided the energy axis into bins and calculated 
the average $\alpha$-helix and $\beta$-strand contents for each bin,
at a fixed temperature near the specific heat maximum.         
Fig.~\ref{fig:5} 
shows the resulting $\alpha$-helix and $\beta$-strand  
profiles $H(E)$ and $S(E)$. We see that the $\beta$-strand content 
$S(E)$ increases steadily with decreasing energy. The $\alpha$-helix 
content $H(E)$, on the other hand, does have its global maximum 
at $E\sim 130$\,kcal/mol. However, the maximum value of $H(E)$ is very 
small. Hence, we find no sign of an obligatory $\alpha$-helical intermediate 
in our model. Most of the amino acids in a typical configuration at 
intermediate energies are either random coils or $\beta$-strands. 
 
\newpage

\subsection{$\beta$-Strand Organization}

As mentioned in the introduction, there exist experimental 
results~\cite{Balbach:00,Gordon:04} suggesting that the 
$\beta$-strands in full \Aba\ fibrils have an in-register, 
antiparallel organization. To find out whether our systems
show a preference for either parallel or antiparallel 
$\beta$-sheets, we consider the joint probability distribution $P(\np,\nm)$,
where $\np$ and $\nm$ count the numbers of interacting chain pairs
with high $\beta$-strand contents 
that are parallel and antiparallel, respectively (see Sec.~\ref{MM}).

Table~\ref{tab:2} shows this distribution for the $\Nc=3$ 
system at $T=275$\,K. For this system, 
the most probable combination of $(\np,\nm)$ is $(1,1)$, corresponding
to a mixed $\beta$-sheet. At the same time, the distribution shows a clear 
asymmetry. The frequency of occurrence for antiparallel
$\beta$-sheets with $(\np,\nm)=(0,2)$ is a factor of 7 higher
than that for parallel $\beta$-sheets with $(\np,\nm)=(2,0)$. 

The corresponding results for $\Nc=6$, at $T=287$\,K, are 
shown in Table~\ref{tab:3}. As in the $\Nc=3$ case, we find that
a majority of the configurations contain mixed $\beta$-sheet structure,
$\np$ and $\nm$ both being nonzero. 
The asymmetry of the $(\np,\nm)$
distribution is  even more pronounced for $\Nc=6$ than for $\Nc=3$.
In particular, we see that large $\nm$ values are much  
more probable than large  $\np$ values; the combination 
$(\np,\nm)=(4,0)$ is, e.g., very unlikely to occur, whereas 
$(\np,\nm)=(0,4)$ does occur with a significant frequency.

Tables~\ref{tab:2} and \ref{tab:3} show the $(\np,\nm)$ distribution at the 
lowest temperatures studied. With increasing temperature, the average 
$\np$ and $\nm$ steadily decrease. 
At the highest temperature studied, 369\,K, about 99\% of the conformations 
have $\np=\nm=0$, for $\Nc=3$ as well as $\Nc=6$. 
The full $(\np,\nm)$ distribution for both $\Nc=3$ and $\Nc=6$ 
at all the different temperatures studied 
can be found as Supplementary Material.

Although the statistical uncertainties are somewhat large, 
the results in Tables~\ref{tab:2} and~\ref{tab:3} show some clear
trends. The most striking one is that large $\np$ values are 
strongly suppressed, which means that large parallel $\beta$-sheets
are very unlikely to form. The probability of having large antiparallel 
$\beta$-sheets is much higher. Compared to purely 
antiparallel $\beta$-sheet structures, it is possible that 
mixed $\beta$-sheet structures are more difficult to extend to large 
stable structures. To be able to check whether or not this is the case, 
simulations of larger systems are required.        

\begin{table}[t]
\begin{center}
\begin{tabular}{rcccc}
$\np$ && \multicolumn{3}{c}{$\nm$}\\
\cline{3-5}
       && 0 & 1 & 2\\
\hline
0 && 0.17\,(2)  & 0.22\,(3)  & 0.14\,(3)  \\
1 && 0.13\,(2)  & 0.32\,(6)  &  \\
2 && 0.020\,(7)  &  &  \\
\end{tabular}
\caption{The probability distribution $P(\np,\nm)$ (see Sec.~\protect\ref{MM})
for $\Nc=3$ \Aba\ peptides at $T=275$\,K. $P(\np,\nm)$ values smaller than
$10^{-3}$ are omitted. The numbers in parentheses are the statistical 
errors in the last digits.} 
\label{tab:2}
\end{center}
\end{table}

\begin{table}[t]
\begin{center}
\begin{tabular}{rcccccc}
$\np$ && \multicolumn{5}{c}{$\nm$}\\
\cline{3-7}
       && 0 & 1 & 2 & 3 & 4 \\
\hline
0 && 0.028\,(5) & 0.059\,(11) & 0.08\,(2) & 0.06\,(2) & 0.030\,(15) \\
1 && 0.038\,(6) & 0.12\,(2) & 0.16\,(3) & 0.10\,(3) & 0.006\,(3)  \\
2 && 0.026\,(11)& 0.11\,(5) & 0.14\,(5) & 0.004\,(2) \\
3 && 0.008\,(5) & 0.013\,(9)  & 0.015\,(12)\\
\end{tabular}
\caption{Same as Table~\protect\ref{tab:2} 
for $\Nc=6$ \Aba\ peptides at $T=287$\,K.} 
\label{tab:3}
\end{center}
\end{table}

Why are antiparallel $\beta$-sheets favored over parallel ones? 
Klimov and Thirumalai~\cite{Klimov:03} concluded that \Aba\ peptides
make antiparallel $\beta$-sheets because of Coulomb interactions 
between charged side chains; the two end side chains of the \Aba\ peptide 
carry opposite charges, which indeed should make the antiparallel 
orientation electrostatically favorable. However, our model completely 
ignores Coulomb interactions between side-chain charges
and still strongly favors the antiparallel organization. 
Other mechanisms than Coulomb interactions between side-chain charges 
might therefore play a significant role, such as the geometry of 
backbone-backbone hydrogen bonds (see Fig.~\ref{fig:1}), steric effects,
and the precise distribution of hydrophobicity along the chains.  
A recent experimental study~\cite{Gordon:04} highlights the 
importance of the hydrophobicity distribution. This study showed that the 
$\beta$-sheet structure of \Aba\ fibrils can be changed from antiparallel 
to parallel by adding an octanyl end group to the peptide which increases 
its amphiphilicity.   

\begin{figure}[t]
\begin{center}
\epsfig{figure=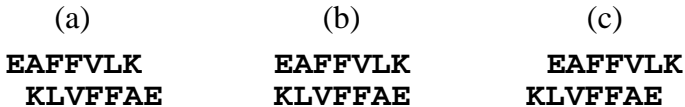,width=9cm}
\end{center}
\vspace{-3mm}
\caption{Schematic representations of three different registries for
an antiparallel pair of \Aba\ peptides.}  
\label{fig:6}\end{figure}

To probe the registry of the $\beta$-sheets, we monitored backbone-backbone 
hydrogen bond patterns (see Fig.~\ref{fig:1}). Fig.~\ref{fig:6} illustrates 
three possible antiparallel registries: (a) $17+k\leftrightarrow 20-k$,
(b) $17+k\leftrightarrow 21-k$, and (c) $17+k\leftrightarrow 22-k$.  
The $17+k\leftrightarrow 21-k$ registry is the one found in experiments on 
\Aba\ fibrils~\cite{Balbach:00,Gordon:04}, whereas experiments on fibrils 
made from the slightly larger segment \Abf\ found evidence for
the $17+k\leftrightarrow 20-k$ registry at pH 7.4 and for the
$17+k\leftrightarrow 22-k$ registry at pH 2.4~\cite{Petkova:04}. 
In our calculations, the $17+k\leftrightarrow 20-k$ and 
$17+k\leftrightarrow 21-k$ registries occur with high and 
comparable frequencies. The $17+k\leftrightarrow 22-k$ registry 
is, by contrast, strongly suppressed, which probably is due 
to hydrophobic effects, although steric clashes between the 
large Phe side chains could play a role, too. 
As to the $17+k\leftrightarrow 20-k$ and $17+k\leftrightarrow 21-k$ 
registries, it would be very interesting to see whether their relative 
frequencies of occurrence depend on $(\np,\nm)$, but that will require 
higher statistics than those provided by the present calculations.

\subsection{Other Peptides} 

To test our model, we performed simulations similar to those for the 
\Aba\ peptide for some other peptides. Some of these peptides, including
the polar one studied by Diaz-Avalos~\etal~\cite{Diaz-Avalos:03},   
had a low overall hydrophobicity. We found that the propensity to aggregate 
is much lower for such peptides than for the \Aba\ peptide, and 
a higher peptide concentration was required to promote aggregation. 
These results clearly show that in our model, hydrophobic attraction 
is a major driving force for aggregation.

As an example of a peptide with a significant hydrophobicity but an 
uneven distribution of it, we studied the peptide 
acetyl-Lys-Phe-Phe-Ala-Ala-Ala-Glu-NH${}_2$, in which the two strongly 
hydrophobic Phe amino acids are asymmetrically placed. For this peptide, 
we obtained aggregated $\beta$-sheet structures with a predominantly 
parallel $\beta$-strand organization, which in particular confirms that 
our model is capable of generating stable parallel $\beta$-sheets.
       
\subsection{Examples of Low-Energy Structures}

It is known that relatively small assemblies formed early in the aggregation  
of full-length A$\beta$~\cite{Lambert:98,Walsh:99,Walsh:02}  as well  as
non-disease-related proteins~\cite{Bucciantini:02} can be toxic, which
makes it very interesting to study possible oligomer shapes. In addition,
such structures represent potential seeds for the fibril formation.  

\begin{figure}[t]
\begin{center}

\epsfig{figure=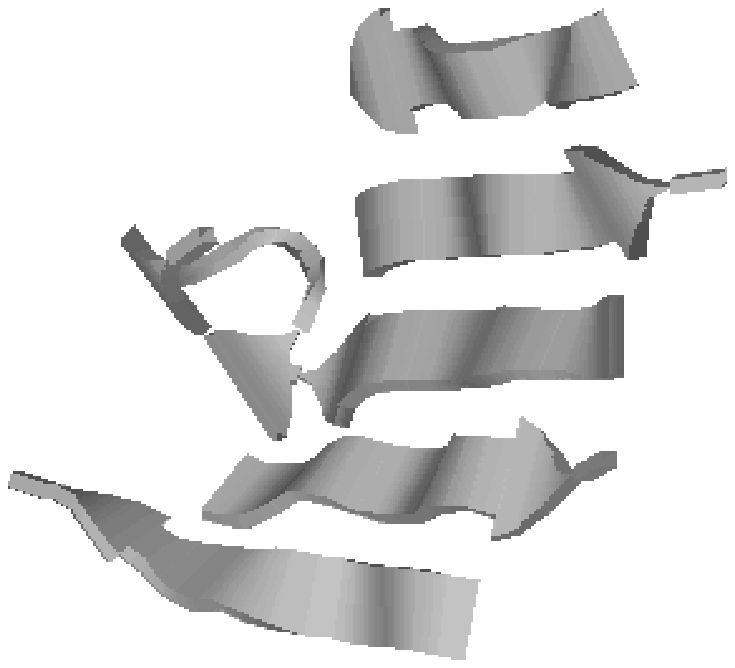,width=4cm}
\hspace{20mm}
\epsfig{figure=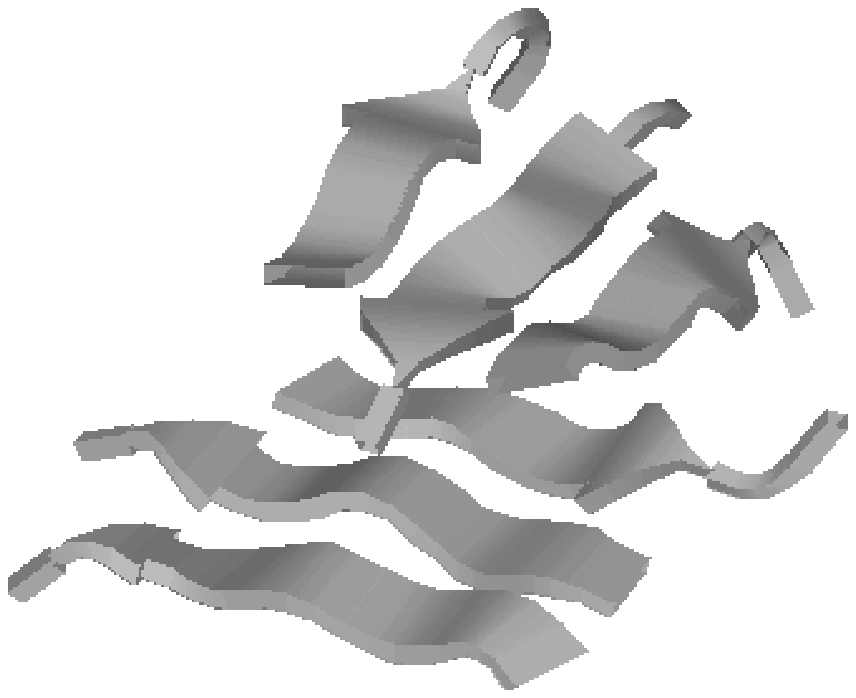,width=4cm}
\end{center}
\vspace{-3mm}
\caption{Two typical low-energy structures from our simulations
of six \Aba\ peptides: 
a five-stranded  $\beta$-sheet (left), and two
three-stranded $\beta$-sheets ``sandwiching'' several of
their  hydrophobic side-chains  between them (right). 
Drawn with RasMol~\cite{Sayle:95}.}
\label{fig:7}\end{figure}

From our simulations, we find that the six-chain \Aba\ system does not exhibit
a single dominating free-energy minimum, but rather a number of more or less
degenerate local minima. Fig.~\ref{fig:7} shows two snapshots of such minima. 
The $\beta$-strand content is, as noted earlier, high, and the 
structures shown in Fig.~\ref{fig:7} illustrate this property.
  
In the simplest class of typical structures observed in our simulations,
five of the chains form a relatively flat $\beta$-sheet, whereas the 
remaining chain is a random coil and held in contact with the $\beta$-sheet 
by hydrophobic attraction. Six-stranded $\beta$-sheets also occur in the
simulations, but with a low frequency, as can be seen from the $P(\np,\nm)$ 
distribution in Table~\ref{tab:3}. Further, for the six-chain system,  
we observe the emergence of new non-trivial structures with no analogs   
in the three-chain simulations. The second structure in Fig.~\ref{fig:7}
illustrates this. Here stability is achieved by stacking two different, 
three-stranded, $\beta$-sheets together, which brings hydrophobic side 
chains from the two $\beta$-sheets in close contact. Such ``sandwiches'' 
occur with a  non-negligible frequency in our simulations.  
To estimate the precise populations of these minima is
difficult. However, five-stranded $\beta$-sheets did occur more
frequently than sandwiches in the simulations. By visual inspection, 
we further estimate that of the order of 10\,\% of the configurations 
are sandwich-like at the 
lowest temperature studied, at which the snapshots were taken. These
low-energy structures also occur at higher temperatures, but become
very rare above the specific heat maximum (see Fig.~\ref{fig:4}).

In none of our simulations did we find any indication of a free-energy 
minimum in which the $\beta$-strands are joined end-to-end to form
the so-called $\beta$-helix~\cite{Wetzel:02}. In our model, 
stability is enhanced  by increasing the number of  hydrogen bonds 
or by increasing hydrophobic contacts. For system  sizes 
as  small as  those  we examined,  the $\beta$-helix  is
inferior  to many  competing structures  in both  these  respects, 
and hence its absence is expected.

\section{Conclusion}

Using a sequence-based atomic model which was originally developed for  
folding studies of single peptides~\cite{Irback:03,Irback:04a,Irback:04b},
we studied the aggregation properties of \Aba\ peptides. In this model, 
we found that \Aba\ peptides have a high propensity to  
self-assemble into aggregated structures with a high $\beta$-strand content,
while the isolated \Aba\ peptide is mainly a random coil. Both parallel 
and antiparallel arrangements of the $\beta$-strands occur in the model, 
with a definite preference for the antiparallel arrangement.

It is important to note that we find this preference for the antiparallel
$\beta$-strand orientation despite ignoring the Coulomb interactions between
the two charged side chains at the ends of the peptide.
It has been suggested~\cite{Klimov:03} that such Coulomb 
interactions are the main determinant for the 
antiparallel orientation. While these Coulomb interactions 
might enhance the tendency for \Aba\ peptides to form $\beta$-sheets 
with an antiparallel organization, our results strongly suggest 
that other factors play a significant role, too. It is worth noting 
that the orientation is not necessarily determined solely by 
sequence-specific side-chain interactions, as antiparallel 
$\beta$-sheets are widely held to be intrinsically more 
stable than parallel ones. For the \Aba\ peptide, which in particular
lacks a clear amphiphility, there is no obvious mechanism 
to overcome this tendency. 

In our simulations, we did not observe an absolute free-energy minimum, 
but rather several nearly degenerate minima corresponding to 
different supra-molecular structures, all consisting of arrangements of 
$\beta$-strands. Apart from single $\beta$-sheets, laminated 
multi-sheet structures were found near free-energy minima for the 
six-chain system. It should be pointed out that the six-chain system
is still too small to permit the formation of, for example, a 
barrel-type structure. It will therefore be very interesting to  
try to extend these calculations to larger system sizes. 

{\large \bf Acknowledgments}

This work was in part supported by the Swedish Research Council 
and the Knut and Alice Wallenberg Foundation through the Swegene consortium.


\newpage

\begin{thebibliography}{}

\bibitem{Rochet:00}
Rochet, J.C., and P.T. Lansbury Jr. 2000.
Amyloid fibrillogenesis: themes and variations.
\COSB 10:\,60--68.

\bibitem{Dobson:03}
Dobson, C.M. 2003.
Protein folding and misfolding.
\Nat 426:\,884--890.

\bibitem{Bucciantini:02}
Bucciantini, M., E. Giannoni, F. Chiti, F. Baroni, L. Formigli, J. Zurdo, 
N. Taddei, G. Ramponi, C.M. Dobson, and M. Stefani. 2002.
Inherent toxicity of aggregates implies a common mechanism
for protein misfolding diseases.
\Nat 416:\,507--511.

\bibitem{Otzen:00}
Otzen, D.E., O. Kristensen, and M. Oliveberg. 2000.
Designed protein tetramer zipped together with a hydrophobic
Alzheimer homology: A structural clue to amyloid assembly.
\PNAS 97:\,9907--9912.

\bibitem{Broome:00}
Broome, B.M., and M.H. Hecht. 2000.
Nature disfavors sequences of alternating polar and non-polar
amino acids: Implications for amyloidogenesis.
\JMB 296:\,961--968. 

\bibitem{Richardson:02}
Richardson, J.S., and D.C. Richardson. 2002.
Natural $\beta$-sheet proteins use negative design to avoid 
edge-to-edge aggregation.
\PNAS 99:\,2754--2759.

\bibitem{West:99}
West, M.W., W.X. Wang, J. Patterson, J.D. Mancias, J.R. Beasley, and 
M.H. Hecht. 1999.
{\it De novo} amyloid proteins from designed combinatorial libraries.
\PNAS 96:\,11211--11216.

\bibitem{Villegas:00}
Villegas, V., J. Zurdo, V.V. Filimonov, F.X. Avil\'es, C.M. Dobson, and
L. Serrano. 2000.
Protein engineering as a strategy to avoid formation of amyloid fibrils.
\ProSci 9:\,1700-1708.

\bibitem{Hammarstrom:02}
Hammarstr\"om, P., X. Jiang, A.R. Hurshman, E.T. Powers, and J.W. Kelly. 2002.
Sequence-dependent denaturation energetics:
a major determinant in amyloid disease diversity. 
\PNAS 99:\,16427--16432.

\bibitem{LopezdelaPaz:02}
L\'opez de la Paz, M., K. Goldie, J. Zurdo, E. Lacroix, C.M. Dobson, 
A. Hoenger, and L. Serrano. 2002.
{\it De novo} designed peptide-based amyloid fibrils.
\PNAS 99:\,16052--16057.  

\bibitem{Chiti:03}
Chiti, F., M. Stefani, N. Taddei, G. Ramponi, and C.M. Dobson. 2003.
Rationalization of the effects of mutations on peptide and protein
aggregation rates.
\Nat 424:\,805--808. 

\bibitem{LopezdelaPaz:04}
L\'opez de la Paz, M., and L. Serrano. 2004.
Sequence determinants of amyloid formation.
\PNAS 101:\,87--92.  

\bibitem{Ventura:04}
Ventura, S., J. Zurdo, S. Narayanan, M. Parre\~no, R. Mangues, B. Reif,
F. Chiti, E. Giannoni, C.M. Dobson, F.X. Aviles, and L. Serrano. 2004.
Short amino acid stretches can mediate amyloid formation in
globular proteins: the Src homology 3 (SH3) case.
\PNAS 101:\,7258--7263.

\bibitem{Sunde:97}
Sunde, M., and C. Blake. 1997.
The structure of amyloid fibrils by electron microscopy and X-ray diffraction.
\APC 50:\,123--159.

\bibitem{Burkoth:00}
Burkoth, T.S., T.L.S. Benzinger, V. Urban, D.M. Morgan, D.M. Gregory, 
P. Thiyagarajan, R.E. Botto, S.C. Meredith, and D.G. Lynn. 2000.
Structure of the $\beta$-amyloid${}_{(10-35)}$ fibril. 
\JACS 122:\,7883--7889.

\bibitem{Petkova:02}
Petkova, A.T., Y. Ishii, J.J. Balbach, O.N. Antzutkin, R.D. Leapman, 
F. Delaglio, and R. Tycko. 2002.
A structural model for Alzheimer's $\beta$-amyloid fibrils based 
on experimental constraints from solid state NMR.
\PNAS 99:\,16742--16747.

\bibitem{Lansbury:95}
Lansbury, P.T., P.R. Costa, J.M. Griffiths, E.J. Simon, M. Auger, 
K.J. Halverson, D.A. Kocisko, Z.S. Hendsch, T.T. Ashburn, R.G.S. Spencer, 
B. Tidor, and R.G. Griffin. 1995.
Structural model for the $\beta$-amyloid fibril based on interstrand alignment 
of an antiparallel-sheet comprising a C-terminal peptide.
\NSB 2:\,990--998.

\bibitem{Petkova:04}
Petkova, A.T., G. Buntkowsky, F. Dyda, R.D. Leapman, W.M. Yau, and
R. Tycko. 2004.
Solid state NMR reveals a pH-dependent antiparallel $\beta$-sheet
registry in fibrils formed by a $\beta$-amyloid peptide.
\JMB 335:\,247--260.

\bibitem{Balbach:00}
Balbach, J.J., Y. Ishii, O.N. Antzutkin, R.D. Leapman, N.W. Rizzo, F. Dyda,
J. Reed, and R. Tycko. 2000.
Amyloid fibril formation by \Aba, a seven-residue fragment of the
Alzheimer's $\beta$-amyloid peptide, and structural characterization
by solid state NMR.
\Bioch 39:\,13748--13759.

\bibitem{Gordon:04}
Gordon, D.J., J.J. Balbach, R. Tycko, and S.C. Meredith. 2004.
Increasing the amphiphilicity of an amyloidogenic peptide changes
the $\beta$-sheet structure in the fibrils from antiparallel to parallel.
\BJ 86:\,428--434.

\bibitem{Tjernberg:96}
Tjernberg, L.O., J. N\"aslund, F. Lindqvist, J. Johansson,
A.R. Karlstr\"om, J. Thyberg, L. Terenius, and C. Nordstedt. 1996. 
Arrest of $\beta$-amyloid fibril formation by a pentapeptide ligand.
\JBC 271:\,8545--8548.

\bibitem{Bratko:01}
Bratko, D., and H.W. Blanch. 2001.
Competition between protein folding and aggregation: a three-dimensional
lattice-model simulation.
\JCP 114:\,561--569.

\bibitem{Harrison:01}
Harrison, P.M., H.S. Chan, S.B. Prusiner, and F.E. Cohen. 2001.
Conformational propagation with prion-like characteristics in a simple model
for protein folding.
\ProSci 10:\,819--835.

\bibitem{Dima:02}
Dima, R.I., and D. Thirumalai. 2002.
Exploring protein aggregation and self-propagation using lattice models:
phase diagram and kinetics.
\ProSci 11:\,1036--1049. 

\bibitem{Jang:04}
Jang, H., C.K. Hall, and Y. Zhou. 2004.
Assembly and kinetic folding pathways of a tetrameric $\beta$-sheet
complex: Molecular dynamics simulations on simplified off-lattice
protein models.
\BJ 86:\,31--49.

\bibitem{Friedel:04}
Friedel, M., and J.E. Shea. 2004.
Self-assembly of peptides into a $\beta$-barrel motif. 
\JCP 120:\,5809--5823.

\bibitem{Ma:02a}
Ma, B., and R. Nussinov. 2002.
Stabilities and conformations of Alzheimer's $\beta$-amyloid 
peptide oligomers (\Aba, \Abb, and \Abc): sequence effects.
\PNAS 99:\,14126--14131.

\bibitem{Ma:02b}
Ma B., and R. Nussinov. 2002.
Molecular dynamics simulations of alanine rich $\beta$-sheet 
oligomers: insight into amyloid formation.
\ProSci 11:\,2335--2350.

\bibitem{Klimov:03}
Klimov, D.K., and D. Thirumalai. 2003.
Dissecting the assembly of \Aba\ amyloid peptides into
antiparallel $\beta$ sheets.
\Str 11:\,295--307.

\bibitem{Gsponer:03}
Gsponer, J., U. Haberth\"ur, and A. Caflisch. 2003.
The role of side-chain interactions in the early
steps of aggregation: Molecular dynamics simulations of an amyloid-forming 
peptide from the yeast prion Sup35.
\PNAS 100:\,5154--5159.

\bibitem{Paci:04}
Paci, E., J. Gsponer, X. Salvatella, and M. Vendruscolo. 2004.
Molecular dynamics studies of the process of amyloid aggregation of 
peptide fragments of transthyrin.
In press: \JMB

\bibitem{Irback:03}
Irb\"ack, A., B. Samuelsson, F. Sjunnesson, and S. Wallin. 2003.
Thermodynamics of $\alpha$- and $\beta$-structure formation in proteins.
\BJ 85:\,1466--1473.
 
\bibitem{Irback:04a}
Irb\"ack, A., and F. Sjunnesson. 2004.
Folding thermodynamics of three $\beta$-sheet peptides: a model study.
\Pro 56:\,110--116.

\bibitem{Irback:04b}
Irb\"ack, A., and S. Mohanty. 2004.
Preprint LU TP 04-28. Submitted to \BJ

\bibitem{Branden:91}
Br\"and\'en, C., and J. Tooze. 1991. 
{\it Introduction to Protein Structure}.
Garland Publishing, New York.

\bibitem{Lyubartsev:92}
Lyubartsev, A.P., A.A. Martsinovski, S.V. Shevkunov, and 
P.N. Vorontsov-Velyaminov. 1992.
New approach to Monte Carlo calculation of the free energy:
method of expanded ensembles,
\JCP 96:\,1776--1783.

\bibitem{Marinari:92}
Marinari, E., and G. Parisi. 1992.
Simulated tempering: A new Monte Carlo scheme,
\EL 19:\,451--458.

\bibitem{Irback:95}
Irb\"ack, A., and F. Potthast. 1995.
Studies of an off-lattice model for protein folding: sequence
dependence and improved sampling at finite temperature,
\JCP 103:\,10298--10305.

\bibitem{Hansmann:99}
Hansmann, U.H.E., and Y. Okamoto. 1999.
New Monte Carlo algorithms for protein folding.
\COSB 9:\,177--183.

\bibitem{Favrin:01}
Favrin, G., A. Irb\"ack, and F. Sjunnesson. 2001.
Monte Carlo update for chain molecules: biased Gaussian steps
in torsional space.
\JCP 114:\,8154--8158.

\bibitem{Ferrenberg:88}
Ferrenberg, A.M., and R.H. Swendsen. 1988.
New Monte Carlo technique for studying phase transitions.
\PRL 61:\,2635--2638. 

\bibitem{Diaz-Avalos:03}
Diaz-Avalos, R., C. Long, E. Fontano, M. Balbirnie, R. Grothe, D. Eisenberg,
and D.L.D. Caspar. 2003.
Cross-beta order and diversity in nanocrystals of an amyloid-forming peptide.
\JMB 330:\,1165--1175.

\bibitem{Lambert:98}
Lambert, M.P., A.K. Barlow, B.A. Chromy, C. Edwards, R. Freed, M. Liosatos,
T.E. Morgan, I. Rozovsky, B. Trommer, K.L. Viola, P. Wals, C. Zhang, 
C.E. Finch, G.A. Krafft, and W.L. Klein. 1998.
Diffusive, nonfibrillar ligands derived from \Abe\ are potent nervous
system neurotoxins.
\PNAS 95:\,6448--6453. 

\bibitem{Walsh:99}
Walsh, D.M., D.M. Hartley, Y. Kusumoto, Y. Fezoui, M.M. Condron, 
A. Lomakin, G.B. Benedek, D.J. Selkoe, and D.B. Teplow. 1999.
Amyloid $\beta$-protein fibrillogenesis --- structure and biological
activity of protofibrillar intermediates.
\JBC 274:\,25945--25952.

\bibitem{Walsh:02}
Walsh, D.M., I. Klyubin, J.V. Fadeeva, W.K. Cullen, R. Anwyl, M.S. Wolfe, 
M.J. Rowan, and D.J. Selkoe. 2002.
Naturally secreted oligomers of amyloid $\beta$ protein potently
inhibit hippocampal long-term potentiation {\it in vivo}.
\Nat 416:\,535--539. 

\bibitem{Sayle:95}
Sayle, R., and E.J. Milner-White. 1995.
RasMol: biomolecular graphics for all.
\TBS 20:\,374--376.

\bibitem{Wetzel:02}
Wetzel, R. 2002.
Ideas of order for amyloid fibril structure.
\Str 10:\,1031--1036.

\end{thebibliography}
\end{document}